\documentclass[aps,pra,superscriptaddress,twocolumn,longbibliography]{revtex4-1}

\usepackage{braket}
\usepackage{float}
\usepackage{graphicx}
\usepackage{amsmath}

\usepackage{units}
\usepackage{amsthm}
\usepackage{amssymb}
\usepackage{graphicx}
\usepackage{color}
\usepackage{bbold}
\usepackage[colorlinks=true, urlcolor=blue, citecolor=blue,linkcolor=blue,citebordercolor={1 0 0},linkbordercolor={0 0 1}]{hyperref}
\usepackage{graphicx}
\graphicspath{{./images/},{./imagesAppendix/}}

\theoremstyle{plain}

\theoremstyle{plain}

\ifx\proof\undefined
\newenvironment{proof}[1][\protect\proofname]{\par
	\normalfont\topsep6\p@\@plus6\p@\relax
	\trivlist
	\itemindent\parindent
	\item[\hskip\labelsep\scshape #1]\ignorespaces
}{%
	\endtrivlist\@endpefalse
}
\providecommand{\proofname}{Proof}
\fi
\theoremstyle{plain}

\theoremstyle{remark}

\makeatother

\newcommand{\idg}[1]{{\bfseries #1)}}

\providecommand{\factname}{Fact}
\providecommand{\theoremname}{Theorem}
\providecommand{\claimname}{Claim}
\providecommand{\lemmaname}{Lemma}
\providecommand{\definitionname}{Definition}

\theoremstyle{definition}

\newcommand{\subfigimg}[3][,]{%
	\setbox1=\hbox{\includegraphics[#1]{#3}}
	\leavevmode\rlap{\usebox1}
	\rlap{\hspace*{2pt}\raisebox{\dimexpr\ht1-0.5\baselineskip}{{\bfseries \large\textsf{#2}}}}
	\phantom{\usebox1}
}

\usepackage{siunitx}
\usepackage{amssymb}
\usepackage{color}
\usepackage{tikz}
\usetikzlibrary{quantikz}
\usepackage{booktabs}

\definecolor{KB}{rgb}{0.4,0.3,0.9}

\definecolor{THc}{rgb}{0.9,0.3,0.2}

\definecolor{JON}{rgb}{0.2,0.8,0.5}

\begin{document}

\title{NISQ Algorithm for Hamiltonian Simulation via Truncated Taylor Series}
\author{Jonathan Wei Zhong Lau}	
\email{e0032323@u.nus.edu}
\affiliation{Centre for Quantum Technologies, National University of Singapore 117543, Singapore}
\author{Tobias Haug}
\affiliation{QOLS, Blackett Laboratory, Imperial College London SW7 2AZ, UK}
\author{Leong Chuan Kwek}
\affiliation{Centre for Quantum Technologies, National University of Singapore 117543, Singapore}
\affiliation{National Institute of Education, Nanyang Technological University, 1 Nanyang Walk, Singapore 637616}
\affiliation{MajuLab, CNRS-UNS-NUS-NTU International Joint Research Unit, UMI 3654, Singapore}
\author{Kishor Bharti}
\email{kishor.bharti1@gmail.com}
\affiliation{Centre for Quantum Technologies, National University of Singapore 117543, Singapore}

\begin{abstract}
Simulating the dynamics of many-body quantum systems is believed to be one of the first fields that quantum computers can show a quantum advantage over classical computers.  
Noisy intermediate-scale quantum (NISQ) algorithms aim at effectively using the currently available quantum hardware. For quantum simulation, various types of NISQ algorithms have been proposed with individual advantages as well as challenges.
In this work, we propose a new algorithm, truncated Taylor quantum simulator (TQS), that shares the advantages of existing algorithms and alleviates some of the shortcomings. Our algorithm does not have any classical-quantum feedback loop and bypasses the barren plateau problem by construction. The classical part in our hybrid quantum-classical algorithm corresponds to a quadratically constrained quadratic program (QCQP)  with a single quadratic equality constraint, which admits a semidefinite relaxation. The QCQP based classical optimization was recently introduced as the classical step in quantum assisted eigensolver (QAE), a NISQ algorithm for the Hamiltonian ground state problem. Thus, our work provides a conceptual unification between the NISQ algorithms for the Hamiltonian ground state problem and the Hamiltonian simulation. We recover differential equation-based NISQ algorithms for Hamiltonian simulation such as quantum assisted simulator (QAS) and variational quantum simulator (VQS) as particular cases of our algorithm. 
We test our algorithm on some toy examples on current cloud quantum computers. We also provide a systematic approach to improve the accuracy of our algorithm.

\end{abstract}

\maketitle

Digital quantum computers have made immense progress in recent years, advancing to solving problems considered to take an unreasonable time to compute for classical computers~\cite{arute2019quantum,zhong2020quantum}. Further, quantum computers are reaching the stage where quantum chemistry problems such as finding the ground state of certain molecules can be achieved within chemical accuracy~\cite{google2020hartree}. 
In short, we are now in the Noisy Intermediate-Scale Quantum (NISQ) era~\cite{preskill2018quantum,bharti2021noisy}, which is characterized by quantum computers with up to a few hundred noisy qubits and lacking full scale quantum error correction.
Thus, noise will limit the usefulness of the computations carried out by these computers~\cite{preskill2018quantum}, preventing algorithms that offer quantum advantage for practical problems, such as Shor's algorithm for prime factorization~\cite{shor1994algorithms}, from being implemented.

However, just because such algorithms cannot be implemented on NISQ devices does not mean that quantum advantage for practical problems cannot be found with NISQ devices. There is currently great interest in the quantum computing and quantum information community to develop algorithms that can be run on NISQ devices but yet deal with problems that are practical~\cite{bharti2021noisy,deutsch2020harnessing,cerezo2020variational}. Some of the most promising avenues deal with the problems in many-body physics and quantum chemistry. One major problem in this field is to develop algorithms capable of estimating the ground state and energy of many-body Hamiltonians. To such ends, algorithms like variational quantum eigensolver (VQE)~\cite{peruzzo2014variational,mcclean2016theory} and quantum assisted eigensolver (QAE) \cite{bharti2020quantum,bharti2020iterative} have been proposed.

The other major problem is to be able to simulate the dynamics of these many-body Hamiltonians. This task can be extremely challenging for classical computers, and Feynman proposed that this would be one of the areas where quantum computers could exhibit clear advantages over classical computers~\cite{feynman1982simulating}. 
Powerful methods to simulate quantum dynamics on fault-tolerant quantum computers have been proposed, such as the concept of using truncated Taylor series by  Berry et al~\cite{berry2015simulating}.

On NISQ devices, a standard approach in simulating quantum dynamics  is to utilize the Trotter-Suzuki decomposition of the unitary time evolution operator into small discrete steps. Each step is made up of efficiently implementable quantum gates, which can be run on the quantum computer~\cite{lloyd1996universal,lanyon2011universal,peng2005quantum,barends2016digitized,barends2015digital,martinez2016real,sieberer2019digital}. However, the depth of the quantum circuit increases linearly with evolution time and the desired target accuracy. On NISQ devices, fidelity rapidly decreases after a few Trotter steps~\cite{poulin2014trotter}, implying long time scales will be unfeasible to simulate with this method.
Alternatives to Trotterization have been proposed, such as VQS~\cite{li2017efficient,yuan2019theory,benedetti2020hardware}, subspace variational quantum simulator (SVQS)~\cite{heya2019subspace}, variational fast forwarding (VFF)~\cite{cirstoiu2020variational,commeau2020variational}, fixed state variational fast forwarding (fsVFF)~\cite{gibbs2021long}, quantum assisted simulator~\cite{bharti2020quantum2,lau2021quantum} and generalized quantum assisted simulator (GQAS)~\cite{haug2020generalized} to name a few.

Recently, Otten, Cortes and Gray have proposed the idea of restarting the dynamics after every timestep by approximating the wavefunction with a variational ansatz~\cite{otten2019noise}. Building on that, Barison, Vicentini and Carleo have proposed a new algorithm~\cite{barison2021efficient} for simulating quantum dynamics. 
Their algorithm, named projected variational quantum dynamics (pVQD) combines the Trotterization and VQS approaches~\cite{li2017efficient,yuan2019theory}. They replace the differential equation with an optimization problem, although not well characterized, and require much simpler circuits compared to VQS. However, pVQD requires a quantum-classical feedback loop and might suffer from the barren plateau problem~\cite{mcclean2018barren} as well the optimization problem may be computationally hard~\cite{bittel2021training}. 
Further, the feedback loop mandates that one has to wait for each computation to finish before the next computation is run, which can be a major bottleneck on cloud-based quantum computers that are accessed via a queue. 

Here, we propose the truncated Taylor quantum simulator (TQS) as new algorithm to simulate quantum dynamics. Our algorithm is building on the ideas of pVQD \cite{otten2019noise,barison2021efficient} combined with the ansatz generation of QAS~\cite{bharti2020quantum2}, which we further enhance by applying the concept of truncated Taylor series by Berry et al~\cite{berry2015simulating}. The contributions of the paper and our algorithm are as following:


\begin{enumerate}
   \item We recast the simulation of the quantum dynamics as a quadratically constrained quadratic program (QCQP). This optimization problem, unlike the optimization problem in pVQD, is well characterized and invites rigorous analysis. The QCQP in our algorithm admits a semidefinite relaxation~\cite{bharti2020quantum}. Moreover, based on ideas from~\cite{bharti2020quantum}, one can provide a sufficient condition for a local minimum to be a global minimum, which a solver can further use as a stopping criterion. Since the classical optimization program in QAE is also a QCQP, it helps us achieve a conceptual unification of TQS with QAE.
    \item The differential equations which form the classical part of QAS and VQS can be recovered in our framework. Since VQS is already a particular case of QAS~\cite{bharti2020quantum2}, our approach yields both VQS and QAS as special cases of TQS.
    \item We remove the need for the classical-quantum feedback loop in pVQD. The absence of the feedback loop yields our algorithm to be exceptionally faster than the feedback loop based NISQ algorithms for simulating quantum dynamics such as~\cite{li2017efficient,heya2019subspace,cirstoiu2020variational,commeau2020variational,gibbs2021long}. The choice of problem-aware ansatz and the structure of the TQS algorithm helps bypass the barren plateau problem.
\end{enumerate}

\medskip
{\noindent {\em TQS Approach---}}
Let us first assume that the Hamiltonian $H$ is expressed as a linear combination of
$r$ tensored Pauli matrices
\begin{gather}
H=\sum_{i=1}^{r}\beta_{i}P_{i}\,,\label{eq:Ham_Pauli}
\end{gather}
with coefficients $\beta_{i}\in\mathbb{C}$. The unitary
evolution under the action of the Hamiltonian $H$ for time $\Delta t$ is given by
\begin{gather}
U\left(\Delta t\right)=\exp\left(-\iota H\Delta t\right)\label{eq:Unitary_1}=\exp\left(-\iota\Delta t\sum_{j=1}^{r}\beta_{j}P_{j}\right)\\
=I-\iota\Delta t\left(\sum_{j=1}^{r}\beta_{j}P_{j}\right)-\frac{\Delta t^{2}}{2}\left(\sum_{j=1}^{r}\beta_{j}P_{j}\right)^{2}+\mathcal{O}\left(\Delta t^{3}\right).\label{eq:Unitary_2}
\end{gather}
We do not need to implement the action of the unitary evolution in such a way. However, for purposes of describing the algorithm for the rest of the paper, we will use this power series expansion first, and talk more about alternatives later. We will now truncate the series, similar to \cite{berry2015simulating}. If we choose small values of $\Delta t$ with respect to the eigen energies of $H$, we can approximate the unitary evolution with $V\left(\Delta t\right)$ 
\begin{gather}
U\left(\Delta t\right)\approx I-\iota\Delta t\left(\sum_{j=1}^{r}\beta_{j}P_{j}\right)\equiv V\left(\Delta t\right).\label{eq:Approx_unitary}
\end{gather}
Let us next choose the ansatz at time $t$ as linear combination of elements
from cumulative $K$-moment states, $\mathbb{CS}_{K}$ (refer to the
QAS paper~\cite{bharti2020quantum2} for the formal definition). These states are defined in the same way as in \cite{bharti2020quantum2} and will be constructed with the help of the given Hamiltonian.
Given a set of $r$ tensored Pauli unitary matrices obtained from the unitary terms of the Hamiltonian $\mathcal{P}\equiv \{P_i\}_{i=1}^r$ and a positive integer $K$ and some efficiently preparable quantum state $\ket{\psi}$, the $K$-moment states are the set of quantum states of the form
\begin{gather}
        \{\ket{\chi_i}\}_i=\{P_{iK}\dots P_{i2}P_{i1}\ket{\psi}\}_{iK,\dots,i2,i1},
\end{gather}
for $P_{il}\in \mathcal{P}$. This set is denoted by $\mathbb{S}_K$. The cumulative $K$-moment states $\mathbb{CS}_K$ are also defined in \cite{bharti2020quantum2} as $\mathbb{CS}_K \equiv \cup _{j=0}^K \mathbb{S}_j$. 

Now the ansatz is expressed as
\begin{gather}
\vert\psi\left(\alpha\left(t\right)\right)\rangle_K=\sum_{\vert\chi_{i}\rangle\in\mathbb{CS}_{K}}\alpha_{i}(t)\vert\chi_{i}\rangle,\label{eq:ansatz_1}
\end{gather}
with some $\alpha_{i}\in\mathbb{C}.$ For small values of $\Delta t$, the
ansatz at time $t+\Delta t$ is given by 
\begin{gather}
\vert\psi\left(\alpha\left(t+\Delta t\right)\right)\rangle_K=\notag \\ \frac{V\left(\Delta t\right)\vert\psi\left(\alpha\left(t\right)\right)\rangle_K}{\left(\langle\psi\left(\alpha\left(t\right)\right)\vert_K V^{\dagger}\left(\Delta t\right)V\left(\Delta t\right)\vert\psi\left(\alpha\left(t\right)\right)\rangle_K\right)^{\frac{1}{2}}}.\label{eq:Evolved_ansatz_1}
\end{gather}
Using the ideas in \cite{barison2021efficient}, our goal now is to variationally approximate the time evolution of the system by adjusting our variational parameters. The crucial difference in our case is that our variational parameters $\alpha$ are coefficients which do not change the basis quantum states $\vert\chi_{i}\rangle$ . Thus, they can be solely updated via a classical computer and do not require a quantum-classical feedback loop. 
To evolve by time $\Delta t$, we update the $\alpha_{i}$ parameters to $\alpha_{i}^{\prime}$
such that the following fidelity measure is maximized
\begin{gather}
F\left(\alpha^{\prime}\right)=\frac{\left|\langle\psi\left(\alpha^{\prime}\right)\vert_K V\left(\Delta t\right)\vert\psi\left(\alpha\right)\rangle_K\right|^{2}}{\langle\psi\left(\alpha\right)\vert_K V^{\dagger}\left(\Delta t\right)V\left(\Delta t\right)\vert\psi\left(\alpha\right)\rangle_K}\label{eq:Fidelity_1}
\end{gather}
Using the notation $\vert\phi\rangle=V\left(\Delta t\right)\vert\psi\left(\alpha\right)\rangle_K,$
the expression for fidelity becomes
\begin{gather}
F\left(\alpha^{\prime}\right)=\frac{\langle\psi\left(\alpha^{\prime}\right)\vert\phi\rangle_K\langle\phi\vert\psi\left(\alpha^{\prime}\right)\rangle_K}{\langle\phi\vert\phi\rangle}.\label{eq:Fidelity_3}
\end{gather}
Using the notation $W_{\phi}\equiv\frac{\vert\phi\rangle\langle\phi\vert}{\braket{\phi| \phi}}$,
the above expression further simplifies to
\begin{gather}
F\left(\alpha^{\prime}\right)=\langle\psi\left(\alpha^{\prime}\right)\vert_K W_{\phi}\vert\psi\left(\alpha^{\prime}\right)\rangle_K.\label{eq:Fdelity_4}
\end{gather}
The goal is to maximize the fidelity subject to the constraint that
$\langle\psi\left(\alpha^{\prime}\right)\vert\psi\left(\alpha^{\prime}\right)\rangle=1.$
Thus, the optimization program at timestep $t$ is given by
\begin{gather}
\max_{\alpha^{\prime}}\text{ }\langle\psi\left(\alpha^{\prime}\right)\vert_K W_{\phi}\vert\psi\left(\alpha^{\prime}\right)\rangle_K
\notag\\
\text{s.t. }\langle\psi\left(\alpha^{\prime}\right)\vert\psi\left(\alpha^{\prime}\right)\rangle_K=1.\label{eq:Optimization_1}
\end{gather}
Using the elements from $\mathbb{CS}_{K}$ and the Hamiltonian $H$,
we define the overlap matrices $\mathcal{E}$ and $\mathcal{D}$ as the following
\begin{gather}
\mathcal{E}_{m,n}=\langle\chi_{m}\vert\chi_{n}\rangle,\label{eq:Overlap_E}\\
\mathcal{D}_{m,n}=\sum_{j}\beta_{j}\langle\chi_{m}\vert P_{j}\vert\chi_{n}\rangle.\label{eq:Overlap_D}
\end{gather}
Because of the way the $\ket{\chi_n}$ states are constructed, these values can be easily computed on a quantum computer, as they simplify to the expectation values of Pauli strings acting on the original quantum state $\ket{\psi}$. 
The constraint in the optimization program \ref{eq:Optimization_1} can written in terms of $\alpha^{\prime}$ as
\begin{gather}
\alpha^{\prime^{\dagger}}\mathcal{E}\alpha^{\prime}=1.\label{eq:Constraiunt_Overlap_1}
\end{gather}
We proceed to write the objective in the optimization program \ref{eq:Optimization_1}
in terms of the overlap matrices $\mathcal{E}$ and $\mathcal{D}$. 
In first order, we can simplify the expression
\begin{gather}
\langle\phi\vert\phi\rangle=\bra{\psi(\alpha)}_K\left(I + (\Delta t)^2 H^2\right)\ket{\psi(\alpha)}_K\notag\\
=\alpha^\dagger \mathcal{E} \alpha + O((\Delta t)^2)\approx \alpha^\dagger \mathcal{E} \alpha. \label{eq:Overlap_obj_2}
\end{gather}
Further, using the notation $G\equiv\left(\mathcal{E}-\iota\Delta t\mathcal{D}\right)$ we find
\begin{gather}
\langle\psi\left(\alpha^{\prime}\right)\vert\phi\rangle_K\langle\phi\vert\psi\left(\alpha^{\prime}\right)\rangle_K=\alpha^{\prime^{\dagger}}G\alpha\alpha^{\dagger}G^{\dagger}\alpha^{\prime}.\label{eq:Overlap_obj_3}
\end{gather}
Using Eq.\ref{eq:Constraiunt_Overlap_1},\ref{eq:Overlap_obj_2},\ref{eq:Overlap_obj_3} and the notation $W_{\alpha}\equiv\frac{G\alpha\alpha^{\dagger}G^{\dagger}}{\alpha^\dagger \mathcal{E} \alpha }$, the optimization program in \ref{eq:Optimization_1}
can be re-expressed in terms of overlap matrices as
\begin{gather}
\max_{\alpha^{\prime}}\text{ }\alpha^{\prime^{\dagger}}W_{\alpha}\alpha^{\prime}\label{eq:Optimization_QCQP_1}\\
\text{s.t }\alpha^{\prime^{\dagger}}\mathcal{E}\alpha^{\prime}=1.\label{eq:Constraint_QCQP_1}
\end{gather}
The aforementioned optimization program is a quadratically constrained quadratic
program with a single equality constraint. As described in \cite{bharti2020quantum}, this QCQP admits a direct convex SDP relaxation. Moreover, the results from~\cite{bharti2020quantum} provide a sufficient condition for a local minimum to be a global minimum, which a solver can further use as a stopping criterion. Alternatively, the problem can be solved with the classic Rayleigh-Ritz procedure by finding the eigenvector associated with the smallest eigenvalue $\lambda$ of the generalized eigenvalue problem $-W_\alpha \alpha^\prime = \lambda E \alpha^\prime$.

It can be shown that in the limit of small $\Delta t$, TQS reduces to QAS (see Appendix \ref{appendix:equivalence}). This could potentially give us a way to obtain systematic higher-order corrections to the QAS matrix differential equation. Interestingly, this is a conceptual unification of the ground state problem (QAE) with the dynamics problem (QAS) in the quantum assisted framework. In QAE, finding the ground state and ground state energy of a Hamiltonian was formulated to become a QCQP. In TQS, the problem of simulating the dynamics is also given as a QCQP. This is conceptually satisfying as the problem of finding the dynamics is  expressed as  $e^{-itH}\ket{\psi}$, which is mathematically similar to using imaginary time evolution to finding the ground state via $e^{-\tau H} \ket{\psi}$. The aforementioned connection is also one of the primary justifications for ansatz selection in \cite{bharti2020iterative}.
We note that as alternative it is possible implement the unitary evolution operator $U(\Delta t)$ directly instead of the Taylor series expansion of Eq.\ref{eq:Evolved_ansatz_1}, however this would require the usage of Hadamard tests (see Appendix \ref{appendix:unitary}).

We want to emphasize again that the quantum computer is only required to measure the overlap matrices $\mathcal{E}$ and $\mathcal{D}$ at the start of the algorithm. No quantum-classical feedback loop for optimization is required. The only optimization steps required are performed solely on the classical computer with knowledge of the overlap matrices. The algorithm is as follows:
    \begin{enumerate}
        \item Choose an efficiently implementable initial state $\ket{\psi}$, then choose some K$>$0 and form the K-moment states $\ket{\chi_i}$ to construct the ansatz. This step can be done on paper.
        \item With knowledge of the Hamiltonian $H$, calculate the overlap matrices $\mathcal{E}$ and $\mathcal{D}$ on the quantum computer. The job of the quantum computer is now done.
        \item Choose a small $\Delta t$ with respect to the eigenvalues of $H$ and evolve the state forward in time using a classical computer, by solving the optimization program \ref{eq:Optimization_QCQP_1} subject to the constraint \ref{eq:Constraint_QCQP_1}. If results are not up to the desired fidelity, increase K and repeat the algorithm.
    \end{enumerate}
The timestep $\Delta t$  could be increased by including higher order terms in the power series expansion of $U(\Delta t)$ in our calculations (Described in Appendix \ref{appendix:higherorder}).

\begin{figure}
    \centering
    \subfigimg[width=0.24\textwidth]{a}{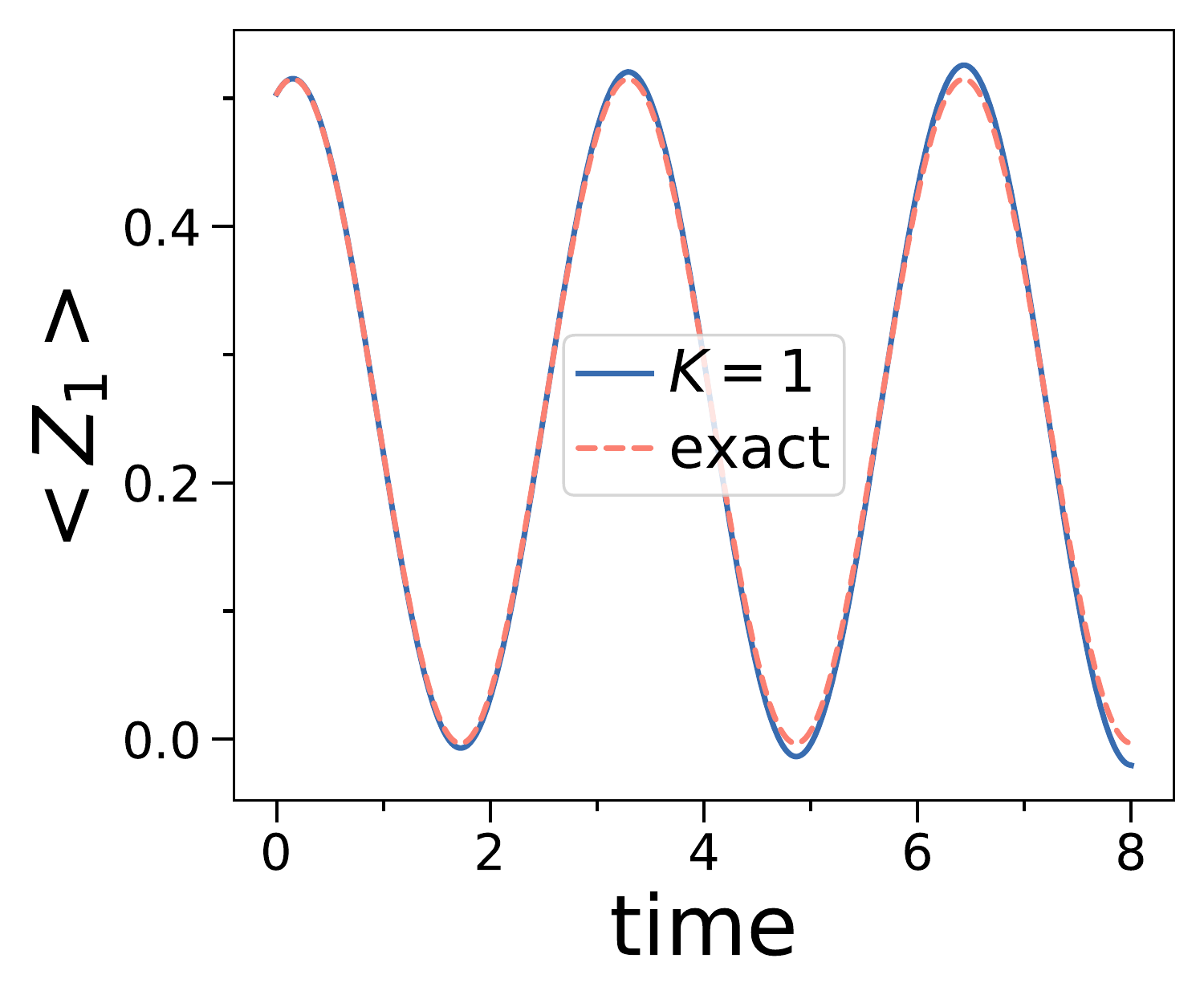}\hfill
    \subfigimg[width=0.24\textwidth]{b}{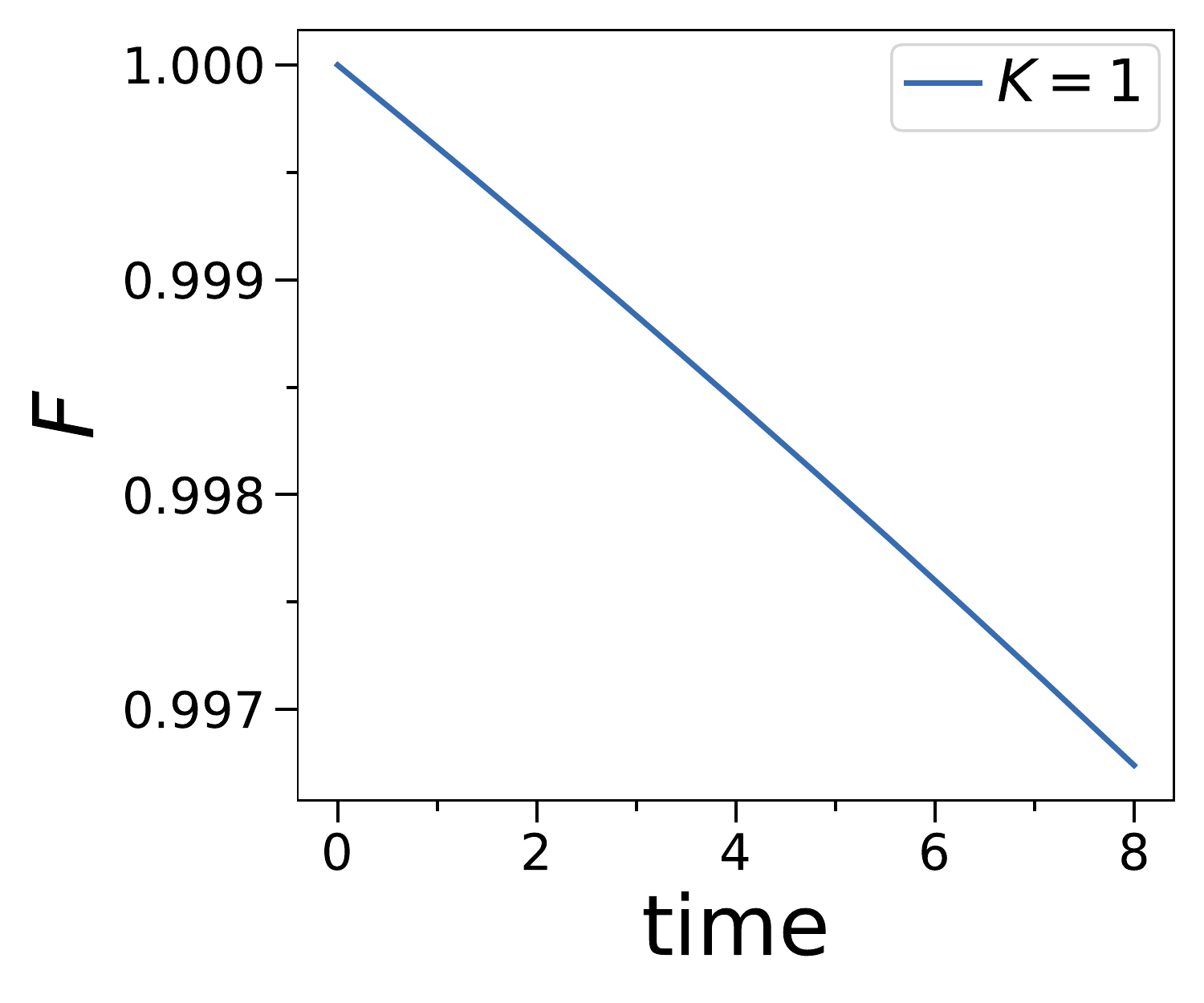}
        \caption{Time evolution of TQS on a 2 qubit state, with Hamiltonian $H_2$, simulated on the IBM quantum processor \emph{ibmq\_rome}. 
        \idg{a} Expectation value of $\braket{Z_1}$
        \idg{b} Fidelity of the state. 
        }
        \label{two_results}
    \end{figure}

\medskip
{\noindent {\em Results---}}
We first use TQS to simulate a 2 qubit Heisenberg model
\begin{gather}
    H_2 = \frac{1}{2}X_1X_2 + \frac{1}{2}Y_1Y_2 + \frac{1}{2}Z_1Z_2.
\end{gather}
We apply it to evolve an initial randomized 2 qubit state $\ket{\psi_2}$.
This initial state is generated by 5 layers of $U_3$ rotations and CNOT gates on the 2 qubits (see Appendix \ref{randomcircuit}). We ran the TQS algorithm on the 5-qubit quantum computer \emph{ibmq\_rome},  available  through  IBM  Quantum  Experience. We used error mitigation by calibrating the measurement errors and applying a filter obtained from that calibration on our data with the toolbox provided in Qiskit~\cite{Qiskit}.
The results are shown in Fig.\ref{two_results}. The evolution of the state under TQS reproduces the exact behavior very well for an ansatz with $K=1$ moment states, even in the presence of the noise of a real quantum computer.

\begin{figure}[htbp]
    \centering
    \subfigimg[width=0.24\textwidth]{a}{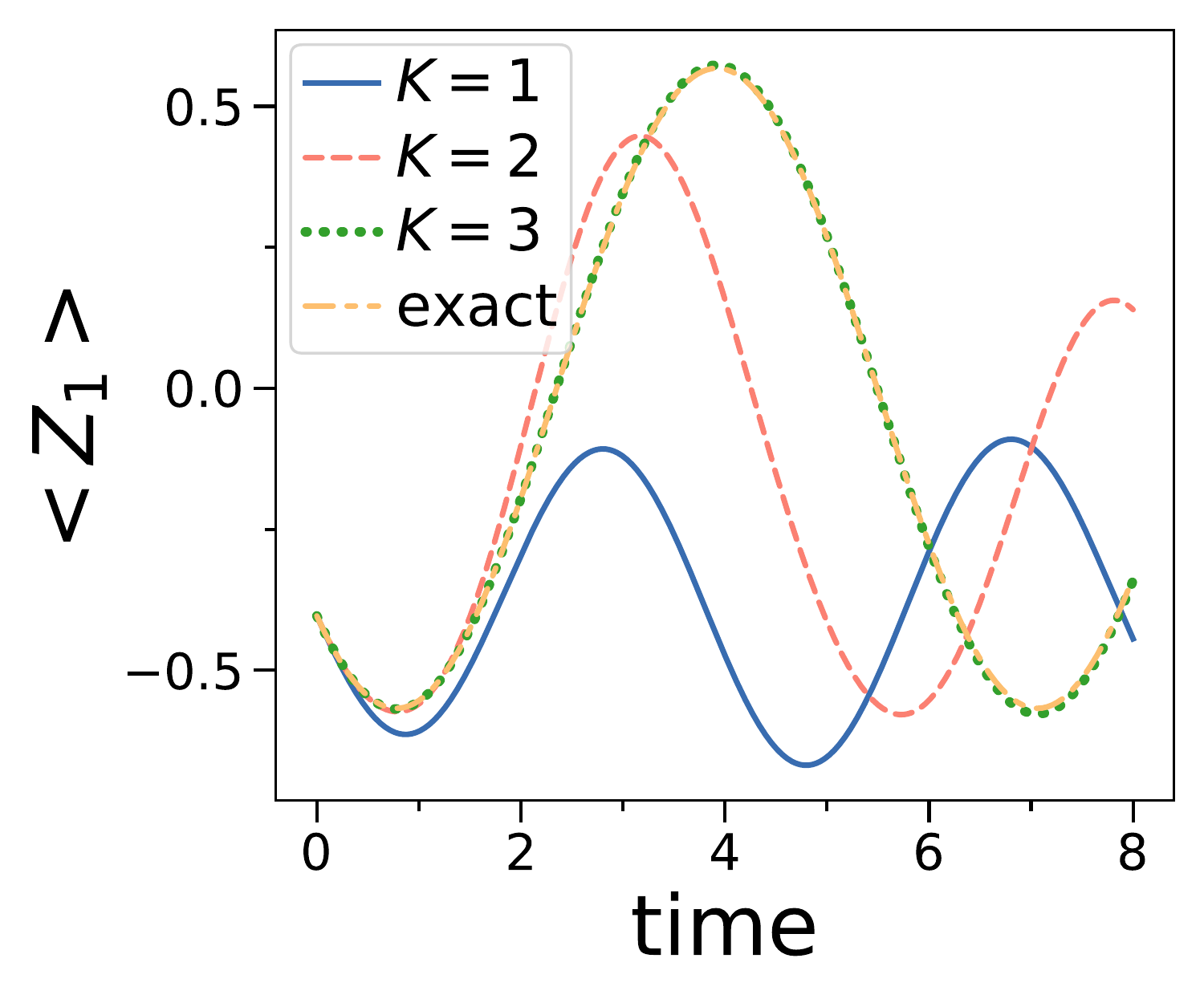}\hfill
    \subfigimg[width=0.24\textwidth]{b}{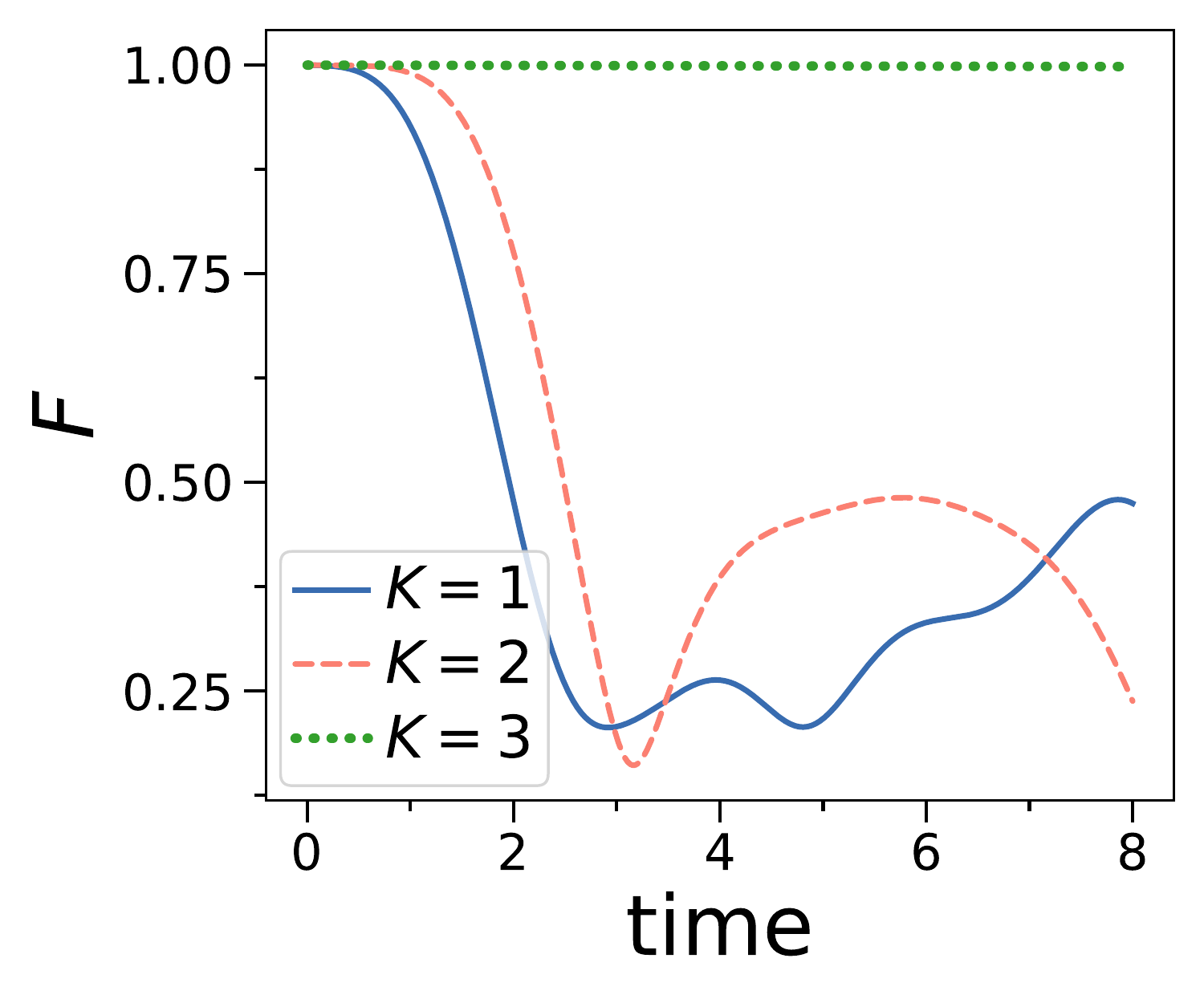}
        \caption{Time evolution of TQS on a 4 qubit state with Hamiltonian $H_4$ simulated on the IBM quantum processor \emph{ibmq\_rome}. 
        \idg{a} Expectation value of $\braket{Z_1}$ 
        \idg{b} Fidelity with exact solution. 
        }
        \label{four_results}
    \end{figure}

Next, we apply TQS to simulate a 4 qubit XX chain model on a quantum computer. Although this Hamiltonian is analytically solvable, we simulate this as a proof of principle.
\begin{gather}
    H_4 = \frac{1}{2}X_1X_2 + \frac{1}{2}X_2X_3 + \frac{1}{2}X_3X_4.
\end{gather}
In Fig.\ref{four_results}, we simulate this Hamiltonian on \emph{ibmq\_rome} with an initial randomized 4 qubit state, generated by 5 layers of $U_3$ rotations and CNOT gates (see Appendix \ref{randomcircuit}). We run it for the $K=1$ to $K=3$ moment states. The evolution of the state under TQS again reproduces the exact behavior very well for the $K=3$ case.

\begin{figure}
    \centering
    \subfigimg[width=0.24\textwidth]{a}{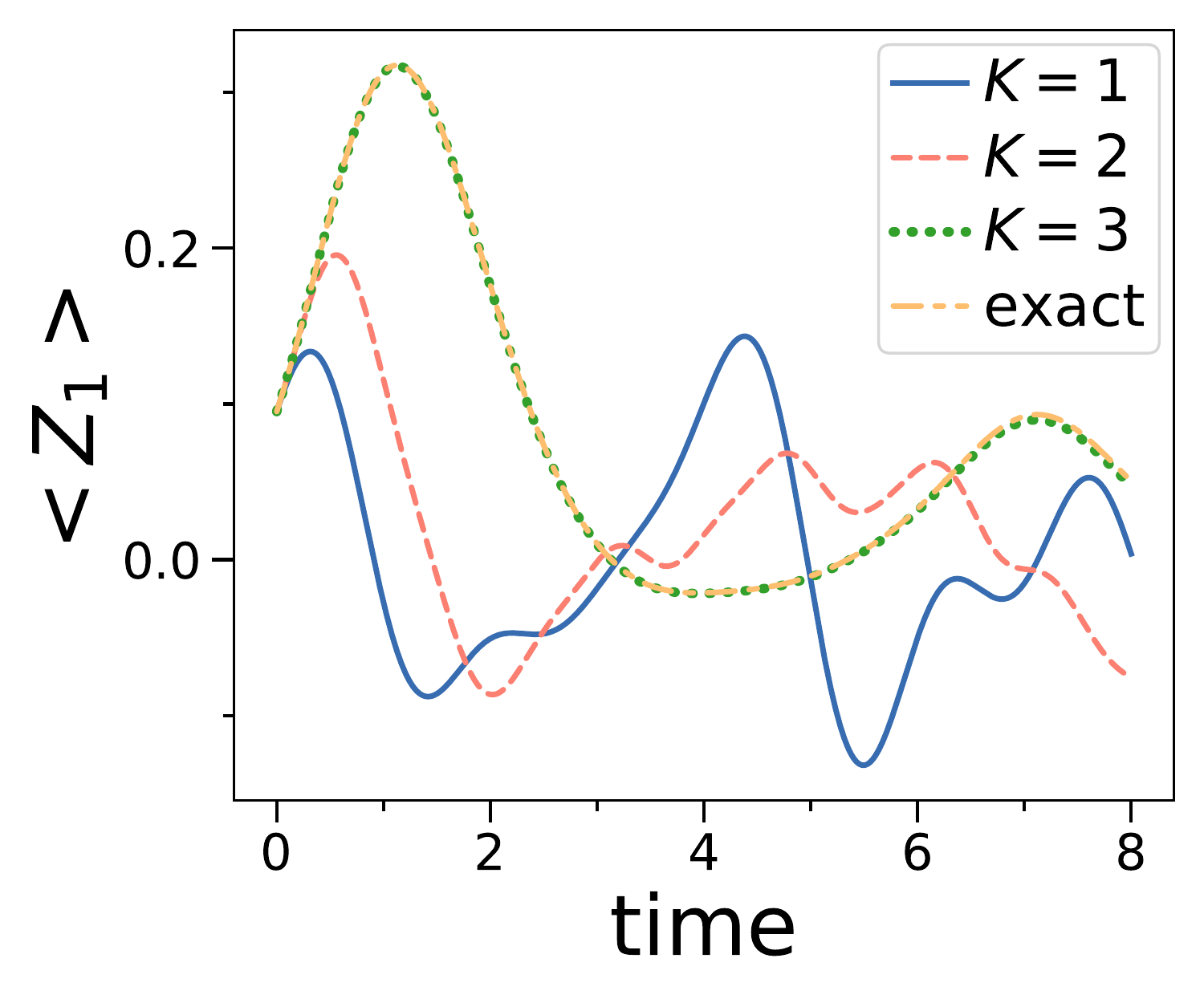}\hfill
    \subfigimg[width=0.24\textwidth]{b}{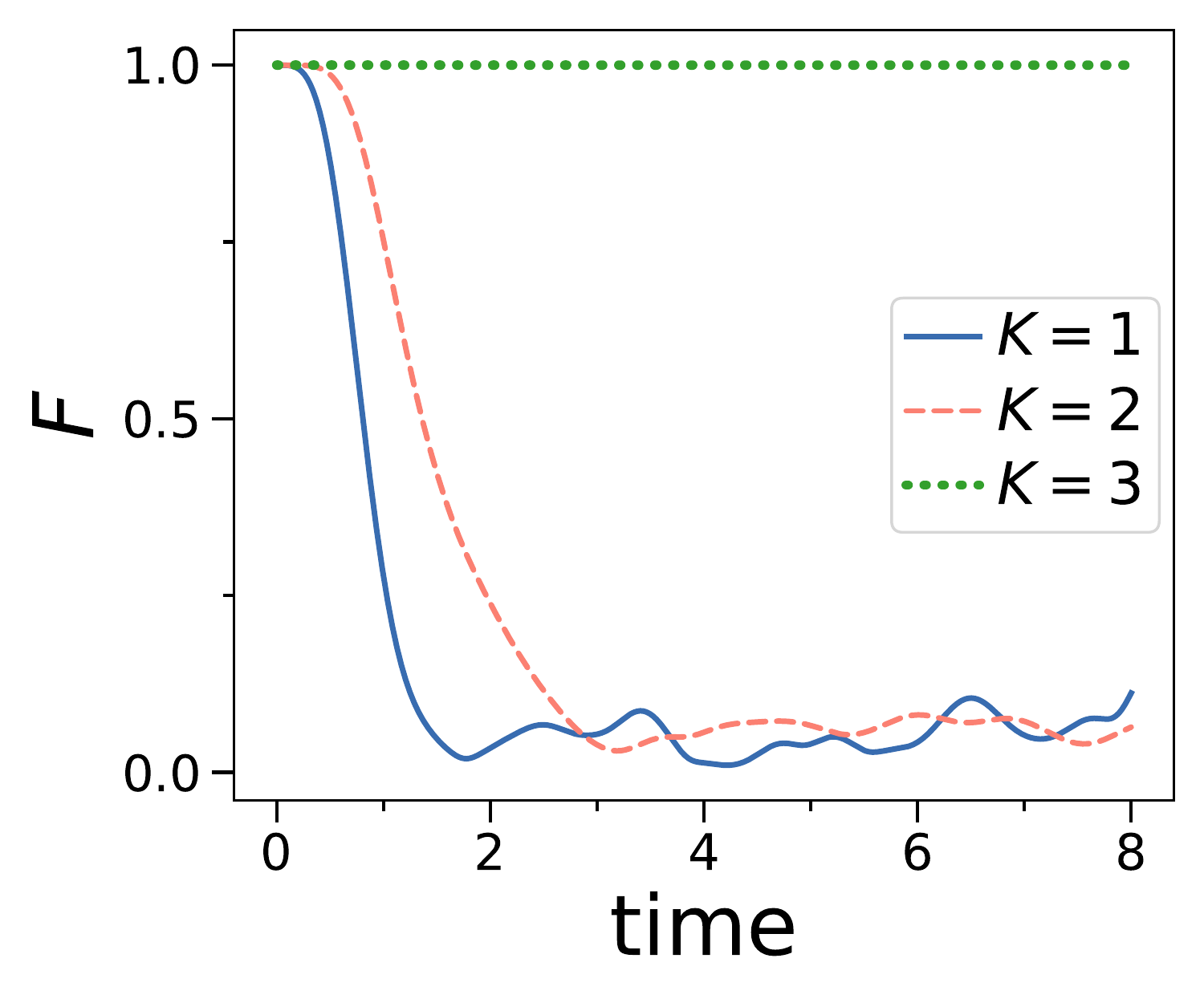}
        \caption{Time evolution of TQS on a 8 qubit state, with Hamiltonian $H_8$, simulated on a classical computer, with a random initial state. \idg{a} Expectation value of $\braket{Z_1}$.
        \idg{b} Fidelity of the state. 
        }
        \label{eight_results}
    \end{figure}

Next, we investigate in Fig.\ref{eight_results} the transverse Ising model with 8 qubits by simulating TQS on a classical computer.
\begin{gather}
    H_8 = \sum_{i=0}^7\frac{1}{2} Z_i Z_{i+1} + \sum_{j=0}^8X_j.
\end{gather}
With an initial random state, we find that the evolution of the state reproduces the exact dynamics for the case of $K=3$ moment expansion. 

\begin{figure}
    \centering
    \includegraphics[width=0.5\textwidth]{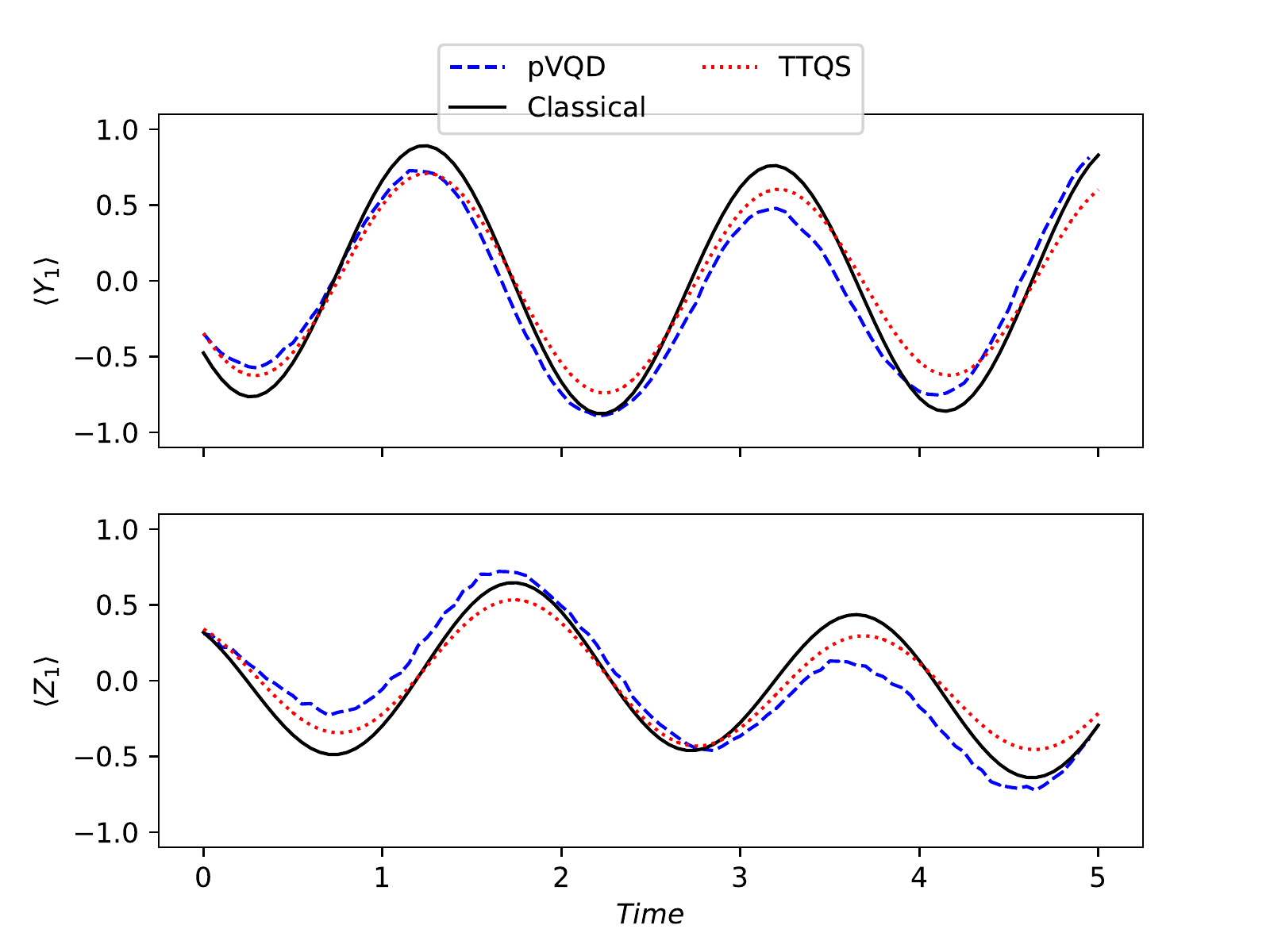}
    \caption{Time evolution of TQS and pVQD on a 2 qubit state, with Hamiltonian $H_{TFI,2}$, simulated with a noisy simulator. The noise model was taken from the IBM Quantum Experience provider, mimicking the noise of the real quantum processor \emph{ibmq\_bogota}. pVQD was run for 100 optimization steps, and made use of a parametric quantum circuit with 8 parameters, made out of sucessive layers of single qubit $X$ rotations and 2-qubit $ZZ$ rotations. The expectation values of $\braket{Y_1}$ and $\braket{Z_1}$ are plotted. The results are somewhat similar, although TQS does have results closer to the classical results most of the time, especially so for the expectation value of $\braket{Z_1}$, as it is better able to capture the peaks and troughs of the expectation values. However, while the results might be argued to be somewhat similar, the resource needs of both algorithms on the quantum computer are quite different. The TQS algorithm in our case required $\approx 30$ circuits to be run, while the pVQD simulator required well over $4000$ circuits to be run.}
    \label{comparison_results}
\end{figure}

Lastly, we compare TQS to pVQD for a 2 qubit transverse Ising model on a simulation. We consider the 2 qubit transverse Ising Hamiltonian:
\begin{gather}
    H_{TFI,2} = \sum_{i=0}^2\frac{1}{2} Z_i Z_{i+1} + \sum_{j=0}^2X_j.
\end{gather}
We compared them with noisy simulators, with the noise models taken from the IBM Quantum Experience provider, which is meant to mimic the noise on their actual quantum computers. The results are shown in Fig.\ref{comparison_results}. As can be seen, while both TQS and pVQD do have errors when trying to simulate this Hamiltonian in the presence of noise, the results for the expectation values of the state for TQS are closer to the classical results most of the time. This is especially so for the expectation value of $\braket{Z_1}$. However, while the results might be argued to be somewhat similar, the resource needs of both algorithms on the quantum computer are quite different. The TQS algorithm in our case required $\approx 30$ circuits to be run, while the pVQD simulator required well over $4000$ circuits to be run, which is already a little prohibitory for us to run on the IBM Quantum Experience. It should be mentioned that if we wanted to increase the simulation time for this example, since the algebra has already closed, we could do that with no extra circuits with TQS, while the number of circuits in pVQD scales linearly with the number of steps required.

\medskip
{\noindent {\em Discussion and Conclusion---}}
The currently proposed NISQ algorithms face problems in scaling up to system sizes where classical computers cannot simulate the same systems, or in other words, to the point where we would see quantum advantage. For example, VQS/SVQS/pVQD require the use of a quantum-classical feedback loop, usually require complicated circuits, share similar problems as VQE like the barren plateau problem, and lack  a systematic way to generate a parameterized ansatz. VFF/fsVFF also suffers from lacking a systematic way to generate the ansatz, usually requires complicated circuits and has to run a quantum-classical feedback loop it at the start. Further, the no fast-forwarding theorem suggests that not all Hamiltonians will be able to be accurately diagonalized with a reasonable amount of gates and circuit length, and the optimization step of the cost function in VFF might be too difficult to carry out efficiently.
However, the barren plateau problem and ansatz state generation could be improved upon by applying various techniques~\cite{haug2021capacity,nakaji2020expressibility,volkoff2020large,holmes2021connecting,huang2019near}.

One problem that VQS and QAS share is that they require solving a differential equation which includes the pseudo-inverse of a matrix, whose elements are measured on a quantum computer. This matrix can be ill-conditioned. This procedure, via singular value decomposition, can be numerically unstable and sensitive to noise, especially as the system increases in size \cite{demmel1987condition}. However, the sensitivity of these matrices has not been rigorously analyzed and more work has to be done to understand the scaling of the sensitivity.

In this work, we develop TQS for simulating quantum dynamics on digital quantum computers. 
TQS recasts the dynamical problem as a QCQP optimization program, which is well characterized unlike the optimization program in pVQD, allowing us to avoid the aforementioned problem in VQS and QAS.

At the same time, TQS retains the advantages of QAS, namely providing us a systematic method to select the ansatz, avoiding complicated Hadamard tests and controlled unitaries, avoiding the barren plateau problem, and only requiring usage of the quantum computer at the start, all of which are problems that are present in pVQD.

However, there are still many problems to tackle in our approach. One problem is an inherited problem from QAS. As the Hamiltonian size and complexity increase, large $K$ values may be needed to generate enough states for a sufficiently expressible ansatz to produce accurate results. Though QAS uses a problem aware ansatz, more information from the problem such as the combination coefficients $\beta_i$ and symmetries of the Hamiltonian should be employed to further tame the complexity.

As the system size increases, it may be required to reduce $\Delta t$ to preserve accuracy in the post-processing part of the algorithm. This will increase the computational cost of the classical computer. The number of optimization steps to be carried out increases linearly with the number of discretizations steps of the evolution time. Determining whether this poses a bottleneck for TQS when applied to large systems requires further studies.

Furthermore, in the presence of noise, the calculated fidelity of our states can go above one. A possible origin are small eigenvalues in the $\mathcal{E}$ overlap matrix, which can give the procedure of optimizing or solving the generalized eigenvalue problem numerical instability. As we scale up the system and consider more ansatz states, this issue can become more prevalent. 

In the future, the NISQ community should investigate these challenges, so that we can successfully run NISQ algorithms for larger qubit numbers.

\medskip
{\noindent {\em Acknowledgements---}}
We are grateful to the National Research Foundation and the Ministry of Education, Singapore for financial support. The authors acknowledge the use of the IBM Quantum Experience devices for this work. This work is supported by a Samsung GRC project and the UK Hub in Quantum Computing and Simulation, part of the UK National Quantum Technologies Programme with funding from UKRI EPSRC grant EP/T001062/1.

\bibliographystyle{apsrev4-1}
\bibliography{TQS}

\newpage
\onecolumngrid
\appendix

\section{Details on running circuits on the IBM quantum computer}\label{randomcircuit}

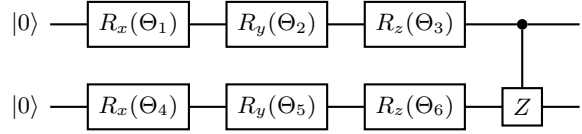
\begin{figure}[H]
    \centering
    \begin{tikzcd}
        \lstick{$\ket{0}$}&\gate{R_x(\Theta_1)}&\gate{R_y(\Theta_2)}&\gate{R_z(\Theta_3)}&\ctrl{1}&\qw& \rstick{}\\
        \lstick{$\ket{0}$}&\gate{R_x(\Theta_4)}&\gate{R_y(\Theta_5)}&\gate{R_z(\Theta_6)}&\gate{Z}&\qw&\rstick{}
    \end{tikzcd}
    \caption{Circuit for two qubits that generate one set of $U_3$ rotation with randomized parameters, followed by a CNOT gate between the 2 qubits. 5 successive layers of this circuit were used to generate the initial starting state for the 2 qubit case on the IBM quantum computer for our runs of TQS. The $\Theta$s were randomly generated.}
    \label{fig:random_circuit1}
\end{figure}

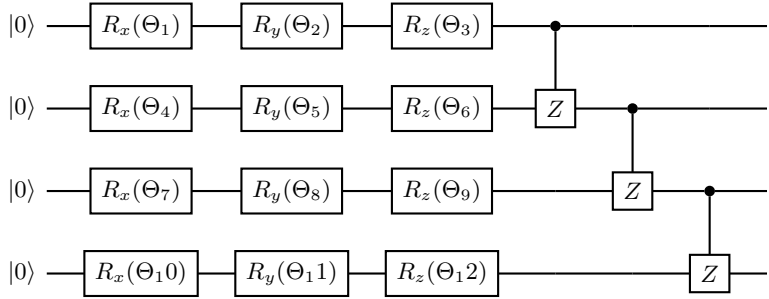
\begin{figure}[H]
    \centering
    \begin{tikzcd}
        \lstick{$\ket{0}$}&\gate{R_x(\Theta_1)}&\gate{R_y(\Theta_2)}&\gate{R_z(\Theta_3)}&\ctrl{1}&\qw&\qw&\qw& \rstick{}\\
        \lstick{$\ket{0}$}&\gate{R_x(\Theta_4)}&\gate{R_y(\Theta_5)}&\gate{R_z(\Theta_6)}&\gate{Z}&\ctrl{1}&\qw&\qw&\rstick{}\\
        \lstick{$\ket{0}$}&\gate{R_x(\Theta_7)}&\gate{R_y(\Theta_8)}&\gate{R_z(\Theta_9)}&\qw&\gate{Z}&\ctrl{1}&\qw&\rstick{}\\
        \lstick{$\ket{0}$}&\gate{R_x(\Theta_10)}&\gate{R_y(\Theta_11)}&\gate{R_z(\Theta_12)}&\qw&\qw&\gate{Z}&\qw&\rstick{}
    \end{tikzcd}
    \caption{Circuit for four qubits that generate one set of $U_3$ rotation with randomized parameters, followed by a series of CNOT gates between the adjacent qubits. 5 successive layers of this circuit were used to generate the initial starting state for the 4 qubit case on the IBM quantum computer for our runs of TQS. The $\Theta$s were randomly generated.}
    \label{fig:random_circuit2}
\end{figure}

For the runs on the real quantum computer, we generated an initial state with randomized parameters to evolve with the following circuit. It comprised 5 layers of successive $U_3$ rotation with randomized parameters on each qubit, followed by a CNOT/entangling gate. We (see Fig.\ref{fig:random_circuit1} and \ref{fig:random_circuit2}). We sampled from each circuit 8192 shots. 

\section{Number of basis states considered for each $K$}\label{appendix:numberstates}

The number of basis states that was used to construct the hybrid ansatz, for each $K$ moment expansion, for each Hamiltonian, is given in Table \ref{tablestates}. 
    \begin{table}[]
    \centering
\begin{tabular}{|l|l|l|l|l|}
\hline
             & $K=1$ & $K=2$ & $K=3$ & $K=4$ \\ \hline
2 Qubit Case & 1     & 4     &       &       \\ \hline
4 Qubit Case & 1     & 4     & 7     & 8     \\ \hline
8 Qubit Case & 1     & 17    &  137     &       \\ \hline
             &       &       &       &       \\ \hline
\end{tabular}
\caption{Comparison of the number of basis states used to construct the hybrid ansatz for each $K$ for each Hamiltonian. For example, the $K=2$ expansion for the 4 qubit case, using the Hamiltonian $H_4$, requires 4 quantum states to construct the hybrid ansatz.}
\label{tablestates}
\end{table}

\section{QAS and VQS as special cases of TQS}\label{appendix:equivalence}

In this appendix, we show that in the limit of choosing a very small $\Delta t$, one obtains QAS from TQS. Since VQS is a special case of QAS \cite{bharti2020quantum2}, we get VQS also as special case of TQS. We start out with the series expansion of $\ket{\psi(\Vec{\alpha} + \delta \Vec{\alpha})}$
\begin{gather}
    \ket{\psi(\Vec{\alpha} + \delta \Vec{\alpha})} = \ket{\psi (\Vec{\alpha})} + \sum_j \frac{\partial}{\partial \alpha_j}\ket{\psi (\Vec{\alpha})} \delta \alpha_j.
\end{gather}
Now in TQS we want to maximize the overlap of $U(\Delta t)\ket{\psi (\Vec{\alpha})}$ and $\ket{\psi (\Vec{\alpha} + \delta \Vec{\alpha})}$, which is essentially the fidelity measure in equation \ref{eq:Fidelity_1}
\begin{gather}
    |\bra{\psi (\Vec{\alpha})}U^\dagger(\Delta t) \ket{\psi (\Vec{\alpha} + \delta \Vec{\alpha})}|^2 = \left[\bra{\psi (\Vec{\alpha})}U^\dagger(\Delta t) \ket{\psi (\Vec{\alpha})} + \sum_j \ket{\psi(\Vec{\alpha})}U^\dagger(\Delta t)\frac{\partial \ket{\psi (\Vec{\alpha})}}{\partial \alpha_j}\delta \alpha_j \right]\times\left[\text{C. C.} \right]\notag \\
    \stackrel{\ket{\psi (\Vec{\alpha})} = \sum_j \alpha_j \ket{\chi_j}}{=}\left[\bra{\psi (\Vec{\alpha})}U^\dagger(\Delta t) \ket{\psi (\Vec{\alpha})} + \sum_j \ket{\psi(\Vec{\alpha})}U^\dagger(\Delta t)\ket{\chi_j}\delta \alpha_j \right]\times\left[\text{C. C.} \right] \notag \\
    = |\bra{\psi (\Vec{\alpha})}U^\dagger(\Delta t) \ket{\psi (\Vec{\alpha})}|^2 + \sum_j \bra{\psi (\Vec{\alpha})}U^\dagger(\Delta t) \ket{\chi_j}\bra{\psi (\Vec{\alpha})}U(\Delta t) \ket{\psi (\Vec{\alpha})}\delta \alpha_j \notag \\
    + \sum_j \bra{\chi_j}U(\Delta t) \ket{\psi (\Vec{\alpha})}\bra{\psi (\Vec{\alpha})}U^\dagger(\Delta t) \ket{\psi (\Vec{\alpha})}\delta \alpha_j^* + \sum_{j,k}\braket{\psi (\Vec{\alpha})|U^\dagger(\Delta t)|\chi_j}\braket{\chi_k|U(\Delta t)|\psi (\Vec{\alpha})}\delta \alpha_j\delta \alpha_k^*.
\end{gather}

Now in the same manner as QAS, using the Mclachlan's variational principle \cite{mclachlan1964variational,yuan2019theory,bharti2020quantum2,lau2021quantum}, we demand that the variation of this fidelity is equal to $0$ with respect to $\alpha_j$:
\begin{gather}
    \implies  \bra{\psi (\Vec{\alpha})}U^\dagger(\Delta t) \ket{\chi_j}\bra{\psi (\Vec{\alpha})}U(\Delta t) \ket{\psi (\Vec{\alpha})} + \sum_{k}\braket{\psi (\Vec{\alpha})|U^\dagger(\Delta t)|\chi_j}\braket{\chi_k|U(\Delta t)|\psi (\Vec{\alpha})}\delta \alpha_k^* = 0 \notag \\
    \implies\bra{\psi (\Vec{\alpha})}U(\Delta t) \ket{\psi (\Vec{\alpha})} + \sum_{k}\braket{\chi_k|U(\Delta t)|\psi (\Vec{\alpha})}\delta \alpha_k^* = 0.
\end{gather}
Now we substitute in $U(\delta t) = I - i \Delta t H$:
\begin{gather}
    \implies \braket{\psi (\Vec{\alpha})\vert\psi (\Vec{\alpha})}-i\Delta t \bra{\psi (\Vec{\alpha})}H\ket{\psi (\Vec{\alpha})} + \sum_{k}\braket{\chi_k|\psi (\Vec{\alpha})}\delta \alpha_k^* - i \Delta t \sum_{k}\braket{\chi_k|H|\psi (\Vec{\alpha})}\delta \alpha_k^* = 0.
\end{gather}
Now we take the derivative of this equation with respect to $\Delta t$. Note that $\frac{d}{d \Delta t}\delta \alpha_k^* = \delta \Dot{\alpha}_k^*$. We then discard any terms remaining that are linear in $\Delta t$ or in $\delta \alpha$ (implying we have chosen such a small $\Delta t$ that $\delta \alpha$ is also very small).
\begin{gather}
    \implies -i \bra{\psi (\Vec{\alpha})}H\ket{\psi (\Vec{\alpha})} + \sum_{k}\delta \Dot{\alpha}_k^*\braket{\chi_k|\psi (\Vec{\alpha})}\delta \alpha_k^* = 0.
\end{gather}
Using the above definition of the $\mathcal{E}$ and $\mathcal{D}$ matrices in equation \ref{eq:Overlap_E} and \ref{eq:Overlap_D}, this simplifies to:
\begin{gather}
    \implies -i \Vec{\alpha}^\dagger \mathcal{D} \Vec{\alpha} + \Vec{\Dot{\alpha}}^\dagger \mathcal{E} \Vec{\alpha} = 0 \notag \\
    \implies \mathcal{E} \Vec{\Dot{\alpha}} = -i \mathcal{D} \Vec{\alpha}.
\end{gather}
This is exactly the same differential equation that we aim to solve in QAS. If we do not ignore the higher order terms, we could obtain systematic higher order corrections to the QAS matrix differential equation using such a method.

\section{Unitary implementation}\label{appendix:unitary}
As alternative, we could implement the unitary evolution operator $U(\Delta t)$ directly instead of the Taylor series expansion of Eq.\ref{eq:Evolved_ansatz_1}
\begin{gather}
\vert\psi\left(\alpha\left(t+\Delta t\right)\right)\rangle_K= U\left(\Delta t\right)\vert\psi\left(\alpha\left(t\right)\right)\rangle_K.
\end{gather}
and defining the matrix $\mathcal{R}_{m,n}=\langle\chi_{m}\vert U(\Delta t) \vert\chi_{n}\rangle$ to solve the program
\begin{gather}
\max_{\alpha^{\prime}}\text{ }\alpha^{\prime^{\dagger}}\mathcal{R}\alpha\alpha^{{\dagger}}\mathcal{R}^\dagger\alpha^{\prime}\label{eq:Optimization_U_QCQP_1}\\
\text{s.t }\alpha^{\prime^{\dagger}}\mathcal{E}\alpha^{\prime}=1\,.\label{eq:Constraint_U_QCQP_1}
\end{gather}
$U(\Delta t)$ could be implemented with a Trotter decomposition or with an oracle. However, this complicates the circuits needed to calculate the $\mathcal{R}$ matrix, requiring the usage of Hadamard tests.

\section{Higher order approximations} \label{appendix:higherorder}
We investigate higher order expansion for the evolution operator in this section.
First, we define the overlap matrix $\mathcal{J}$ 
\begin{gather}
    \mathcal{J}_{m,n}=\sum_{i,j}\beta_{i}\beta_{j}\langle\chi_{m}\vert P_{i} P_{j}\vert\chi_{n}\rangle.\label{eq:Overlap_J}
\end{gather}
Considering the next highest power expansion of $U(\Delta t)$:
\begin{gather}
    U(\Delta t) \approx I-\iota\Delta t\left(\sum_{j=1}^{r}\beta_{j}P_{j}\right) - \frac{\Delta t^2}{2}\left(\sum_{j=1}^{r}\beta_{j}P_{j}\right)^2\equiv V_2\left(\Delta t\right), 
\end{gather}
and defining $\ket{\phi}=V_2(\Delta t) \ket{\psi(\alpha)}_K$, the constraint in the optimization program \ref{eq:Optimization_1} turns out to be still the same as equation \ref{eq:Constraiunt_Overlap_1}:
\begin{gather}
\braket{\psi(\alpha^{\prime^\dagger})|\psi(\alpha^{\prime^\dagger})}=\alpha^{\prime^{\dagger}}\mathcal{E}\alpha^{\prime} .\label{eq:Overlap2_obj_1}
\end{gather}
It turns out that $\langle\phi\vert\phi\rangle$ is actually exactly equal to $ \alpha^\dagger E \alpha$, which is the result we used earlier in equation \ref{eq:Overlap_obj_2}, as all the 2nd order terms nicely cancel out.

Now, using the notation $G_2\equiv\left(\mathcal{E}-\iota\Delta t \mathcal{D} -\frac{\Delta t^2}{2}\mathcal{J}\right)$,
\begin{gather}
\langle\psi\left(\alpha^{\prime}\right)\vert\phi\rangle_K\langle\phi\vert\psi\left(\alpha^{\prime}\right)\rangle_K=\alpha^{\prime^{\dagger}}G_2\alpha\alpha^{\dagger}G_2^{\dagger}\alpha^{\prime}.\label{eq:Overlap2_obj_3}
\end{gather}

Now the optimization program in \ref{eq:Optimization_1}
can be re-expression in this higher order approximation as

\begin{gather}
\max_{\alpha^{\prime}}\text{ }\alpha^{\prime^{\dagger}}\left(\frac{G_2\alpha\alpha^{\dagger}G_2^{\dagger}}{\alpha^\dagger \mathcal{E} \alpha }\right)\alpha^{\prime}\label{eq:Optimization2_2}
\end{gather}
\[
\text{s.t }\alpha^{\prime^{\dagger}}\mathcal{E}\alpha^{\prime}=1.
\]
And using the notation $W_{2,\alpha}\equiv\frac{G_2\alpha\alpha^{\dagger}G_2^{\dagger}}{\alpha^\prime \mathcal{E} \alpha},$
we further condense the above optimization program as
\begin{gather}
\max_{\alpha^{\prime}}\text{ }\alpha^{\prime^{\dagger}}W_{2,\alpha}\alpha^{\prime}\label{eq:Optimization2_QCQP_1}\\
\text{s.t }\alpha^{\prime^{\dagger}}\mathcal{E}\alpha^{\prime}=1.\label{eq:Constraint2_QCQP_1}
\end{gather}
Once again, the only work that the quantum computer need to do is to calculate overlap matrices in the start, in this case having to calculate $\mathcal{E}$, $\mathcal{D}$ and $\mathcal{J}$. In fact, when going from lower order approximations to higher order approximations, you can reuse the saved matrices and only calculate the new ones needed. In this case, in the original TQS, which uses a first order approximation for $U(\Delta t)$, we already have the $\mathcal{E}$ and $\mathcal{D}$ matrices, so if we deem the results not up to our desired accuracy, we can easily go to the second order approximation showed here, and only require calculation of one additional matrix $\mathcal{J}$.

\end{document}